\documentclass[letterpaper,twocolumn,prl,aps,superscriptaddress,floatfix]{revtex4}

\usepackage[latin9]{inputenc}
\usepackage{amsmath}
\usepackage{amssymb}
\usepackage{graphicx}
\usepackage{esint}
\usepackage{times}

\usepackage[unicode=true,
bookmarks=true,bookmarksnumbered=false,bookmarksopen=false,
breaklinks=false,pdfborder={0 0 1},backref=false,colorlinks=true]
{hyperref}
\hypersetup{
    linkcolor=magenta, urlcolor=blue, citecolor=blue, pdfstartview={FitH}, hyperfootnotes=false, unicode=true}

\makeatletter

\pdfpageheight\paperheight
\pdfpagewidth\paperwidth

\@ifundefined{textcolor}{}
{%
    \definecolor{BLACK}{gray}{0}
    \definecolor{WHITE}{gray}{1}
    \definecolor{RED}{rgb}{1,0,0}
    \definecolor{GREEN}{rgb}{0,1,0}
    \definecolor{BLUE}{rgb}{0,0,1}
    \definecolor{CYAN}{cmyk}{1,0,0,0}
    \definecolor{MAGENTA}{cmyk}{0,1,0,0}
    \definecolor{YELLOW}{cmyk}{0,0,1,0}
}

\usepackage{xcolor}
\usepackage{soul}

\newcommand{\bra}[1]{\ensuremath{\left\langle#1\right|}}
\newcommand{\ket}[1]{\ensuremath{\left|#1\right\rangle}}
\definecolor{blue}{rgb}{0,0,1}
\definecolor{red}{rgb}{1,0,0}
\definecolor{green}{rgb}{0,1,0}

\makeatother

\begin{document}

\title{Scalable Method for Eliminating Residual $ZZ$ Interaction between Superconducting Qubits}

\author{Zhongchu Ni}
\thanks{These authors contributed equally to this work.}
\affiliation{Shenzhen Institute for Quantum Science and Engineering, Southern University of Science and Technology, Shenzhen 518055, China}
\affiliation{International Quantum Academy, Shenzhen 518048, China}
\affiliation{Department of Physics, Southern University of Science and Technology, Shenzhen 518055, China}
\affiliation{Guangdong Provincial Key Laboratory of Quantum Science and Engineering, Southern University of Science and Technology, Shenzhen 518055, China}

\author{Sai Li}
\thanks{These authors contributed equally to this work.}

\author{Libo Zhang}
\affiliation{Shenzhen Institute for Quantum Science and Engineering, Southern University of Science and Technology, Shenzhen 518055, China}
\affiliation{International Quantum Academy, Shenzhen 518048, China}
\affiliation{Guangdong Provincial Key Laboratory of Quantum Science and Engineering, Southern University of Science and Technology, Shenzhen 518055, China}

\author{Ji Chu}

\author{Jingjing Niu}
\author{Tongxing Yan}
\author{Xiuhao Deng}
\author{Ling Hu}
\author{Jian Li}
\author{Youpeng Zhong}
\author{Song Liu}

\author{Fei Yan}
\email{yanf7@sustech.edu.cn}

\author{Yuan Xu}
\email{xuy5@sustech.edu.cn}
\affiliation{Shenzhen Institute for Quantum Science and Engineering, Southern University of Science and Technology, Shenzhen 518055, China}
\affiliation{International Quantum Academy, Shenzhen 518048, China}
\affiliation{Guangdong Provincial Key Laboratory of Quantum Science and Engineering, Southern University of Science and Technology, Shenzhen 518055, China}

\author{Dapeng Yu}
\affiliation{Shenzhen Institute for Quantum Science and Engineering, Southern University of Science and Technology, Shenzhen 518055, China}
\affiliation{International Quantum Academy, Shenzhen 518048, China}
\affiliation{Department of Physics, Southern University of Science and Technology, Shenzhen 518055, China}
\affiliation{Guangdong Provincial Key Laboratory of Quantum Science and Engineering, Southern University of Science and Technology, Shenzhen 518055, China}

\begin{abstract}
Unwanted $ZZ$ interaction is a quantum-mechanical crosstalk phenomenon which correlates qubit dynamics and is ubiquitous in superconducting qubit systems. It adversely affects the quality of quantum operations and can be detrimental in scalable quantum information processing. Here we propose and experimentally demonstrate a practically extensible approach for complete cancellation of residual $ZZ$ interaction between fixed-frequency transmon qubits, which are known for long coherence and simple control. We apply to the intermediate coupler that connects the qubits a weak microwave drive at a properly chosen frequency in order to noninvasively induce an ac Stark shift for $ZZ$ cancellation. We verify the cancellation performance by measuring vanishing two-qubit entangling phases and $ZZ$ correlations. In addition, we implement a randomized benchmarking experiment to extract the idling gate fidelity which shows good agreement with the coherence limit, demonstrating the effectiveness of $ZZ$ cancellation. Our method allows independent addressability of each qubit-qubit connection, and is applicable to both nontunable and tunable couplers, promising better compatibility with future large-scale quantum processors. 
\end{abstract}
\maketitle
\vskip 0.5cm

Scalable quantum information processing relies on simultaneous implementation of high-fidelity quantum operations. Unfortunately, unwanted crosstalk effects become inevitable in a quantum processor with a high degree of integration, compromising quantum operational fidelity and ultimately limiting scalability.
Recent progress with superconducting qubits has shown that by leveraging a tunable-coupling architecture, various crosstalk phenomena can be efficiently reduced while fast two-qubit gates are enabled~\cite{Yan2018,Arute2019}. However, residual longitudinal or $ZZ$ interaction -- by which the frequency of a qubit depends on the state of the other -- may still exist due to the fact that superconducting qubits are not so well-defined two-level systems by nature and the computational states are more or less affected by noncomputational states that are energetically close.
Such unwanted $ZZ$ interaction will result in spectator errors~\cite{Sundaresan2020, Krinner2020, Cai2021, Zajac2021}, correlated errors~\cite{Postler2018, Uwe2020}, and coherent phase errors during intermittent idling operations~\cite{Barends2016, Kandala2019, Gong2019, Karamlou2021}.
Notably, quantum error correction usually necessitates relatively lengthy reset and feedback operations, making it extremely susceptible to these coherent errors accrued during idling~\cite{huang2020, Bultink2020, Andersen2020, Chen2021}. 

There are several methods for $ZZ$ suppression or cancellation, such as, among the passive ones, the use of large qubit-qubit detuning~\cite{DiCarlo2009, Collodo2020, Xu2020}, qubits with opposite anharmonicity signs~\cite{Zhao2020, Ku2020, XuXuexin2021}, and multiple coupling paths~\cite{Mundada2019, Li2020, kandala2021, Zhao2021, Sete2021Parametric, Stehlik2021, XuXuexin2022}.
These methods either result in incomplete cancellation or heavily rely on design and fabrication precision which can be hard to achieve in practice.
There are also a few active methods. One approach is to apply dynamical-decoupling sequences during long idling periods at the expense of additional gates~\cite{Jurcevic2021, Tripathi2021}. It is, however, unclear how to best implement dynamical-decoupling protocols in the context of a highly connected qubit array.   
Another approach, which utilizes the ac Stark effect by off-resonantly driving the qubits, has been demonstrated in recent experiments at a small scale~\cite{Atsushi2020,mitchell2021,wei2021,xiong2021}. Although suitable for architecture with fixed-frequency qubits, the Stark method requires independent drives applied to at least one -- sometimes both -- of the two qubits in order to cancel the $ZZ$ coupling between this particular qubit pair. In a high-connectivity qubit array such as two-dimensional grid where the number of nearest-neighbor connections is more than the number of qubits, this method becomes inextensible due to the lack of degrees of freedom in control.
In addition, directly driving the qubit may induce spurious excitation and leakage to the target qubit and its surroundings, especially in a frequency-crowded system.

In this Letter, we propose and experimentally demonstrate a practically scalable and noninvasive method for cancelling residual $ZZ$ interaction between fixed-frequency superconducting qubits. In contrast to previous approaches that utilize ac Stark shift by driving the qubits, we apply a weak microwave drive to the intermediate coupler instead. On a device with two transmon qubits connected to a cavity bus coupler, we show complete removal of residual $ZZ$ interaction from measurement of two-qubit entangling phases and $ZZ$ correlations. We also perform simultaneous randomized benchmarking~(RB) experiments with interleaved idling operations to extract idling gate fidelity which shows full-scale improvement, reaching the coherence limit.

\begin{figure}
    \includegraphics{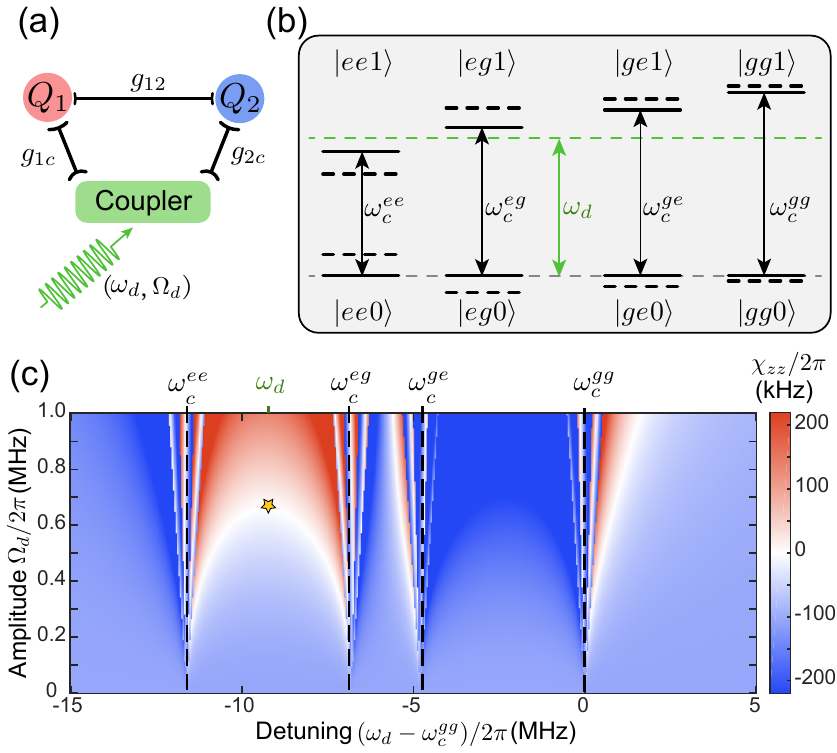}
    \caption{Coupler-assisted $ZZ$ cancellation. 
    \textbf{(a)} Schematic of the system consisting tof wo qubits and an intermediate coupler. The $ZZ$ cancellation drive of frequency $\omega_d$ and amplitude $\Omega_d$ is applied to the coupler. 
    \textbf{(b)} Diagram of the relevant energy levels denoted by $\ket{Q_1,Q_2,C}$. The solid (dashed) black lines represent the corresponding energy levels without (with) externally applied drive to the coupler (dashed green line). Arrows indicate corresponding frequencies.
    \textbf{(c)} Numerical simulated net $ZZ$ coupling $\chi_{zz}$ as a function of the drive frequency $\omega_d$ referenced to $\omega_c^{gg}$ and the drive amplitude $\Omega_d$. Dashed lines indicate the state-dependent coupler frequencies $\omega_c^{mn}$. The star marks the optimal choice of drive parameters for complete $ZZ$ cancellation.}
    \label{fig1}
\end{figure}

Consider a general model described in Fig.~\ref{fig1}(a), where two qubits $Q_1$ and $Q_2$ couple to an intermediate coupler $C$ with a coupling strength of $g_{1c}$ and $g_{2c}$, respectively, as well as to each other with a coupling strength $g_{12}$. 
The static Hamiltonian can be written as ($\hbar = 1$):
\begin{eqnarray}
\label{H1}
H_0 &=& \sum_{i=1,2,c}{\omega_i \, a_i^\dag a_i + \frac{\eta_i}{2} \, a_i^\dag a_i^\dag a_i a_i} \notag\\
&+&  \sum_{i\neq j}{g_{ij} \left( a_i^\dag a_j + a_i a_j^\dag \right)},
\end{eqnarray}
where $\omega_i$ ($i\in \{ 1, 2, c \}$) is the bare frequency of the qubits or coupler, $\eta_i$ is the anharmonicity of each mode, and $a_i$ ($a_i^\dag$) is the corresponding annihilation (creation) operator.
In the dispersive limit, where $g_{ij} \ll \left| \omega_i - \omega_j \right|$, we can conveniently diagonalize the Hamiltonian and truncate it to the relevant subspace:
\begin{eqnarray}
\label{H2}
H_0 &=& \sum_{m,n=g,e}{\ket{mn}\bra{mn} \otimes \left(\omega_c^{mn} a_c^\dag a_c + \frac{\eta_c}{2} \, a_c^\dag a_c^\dag a_c a_c\right)}  \notag\\
&+ &  \chi_{zz}^\mathrm{static}\,\ket{ee}\bra{ee}\otimes\ket{0}\bra{0},
\end{eqnarray}
which is expressed in the rotating frame of the qubits.
For clarity, we use $\ket{g}$ ($\ket{e}$) to denote the ground (excited) state of the qubits and $\ket{0}$, $\ket{1}$, $\ket{2}$... to denote the coupler state. 
$\omega_c^{mn}$ is the 0-1 transition frequency of the coupler with the two qubits in state $\ket{mn}$. In general, the four values of $\omega_c^{mn}$ are different as illustrated in Fig.~\ref{fig1}(b). We shall later exploit such inhomogeneity for $ZZ$ cancellation. 
$\chi_{zz}^\mathrm{static}$ is the residual $ZZ$ coupling, which results from a finite effective coupling between the two qubits and causes unwanted $ZZ$ interaction during idling periods. Detailed discussions about $\chi_{zz}^\mathrm{static}$ can be found in Ref.~\cite{Chu2021}.

\begin{figure}
    \includegraphics{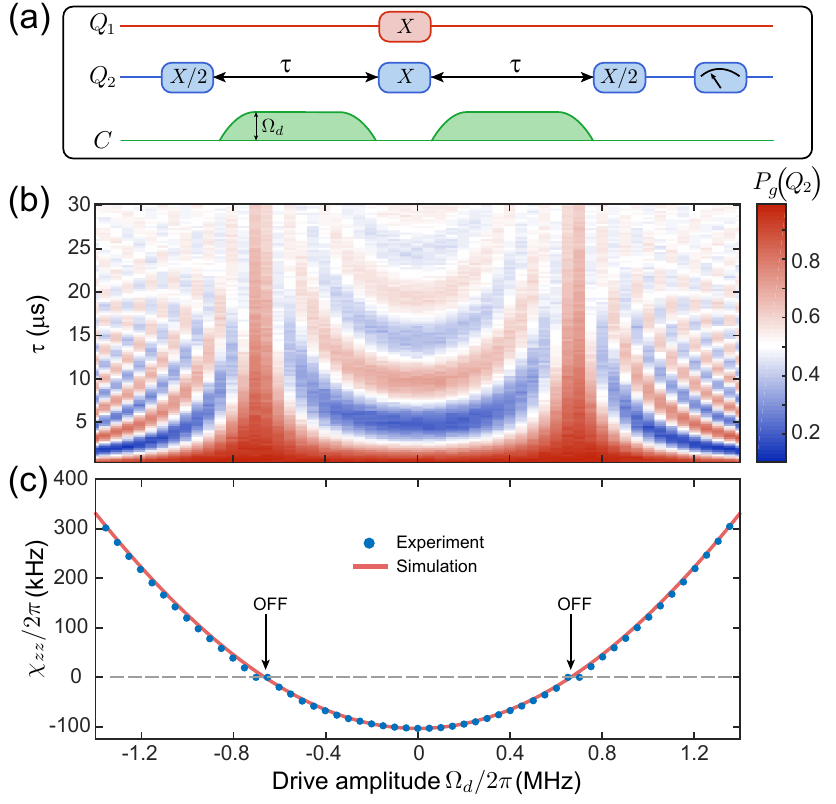}
    \caption{Tunable $ZZ$ coupling $\chi_{zz}$. 
    \textbf{(a)} Pulse sequence to measure the $ZZ$ coupling.
    During the two idling periods of duration $\tau$, a microwave pulse with a same length (including 300~ns rise and fall) and a variable amplitude $\Omega_d$ is applied to the coupler for tuning $ZZ$ coupling. 
    \textbf{(b)} Measured $Q_2$ population as a function of $\tau$ and $\Omega_d$. The fringe pattern indicates the accrued $ZZ$ phase. 
    \textbf{(c)} Extracted $\chi_{zz}$ (dots) from fitting oscillations in (b). The sign is deduced from theory. The \lq\lq{OFF}\rq\rq{} signs indicate where $\chi_{zz}=0$. Solid line is from numerical simulation.}
    \label{fig2}
\end{figure}

To eliminate $\chi_{zz}^\mathrm{static}$ from the system, we apply a microwave drive $\frac{\Omega_d}{2} \left(a_c^\dag \textrm{e}^{-\textrm{i}\omega_d t} + a_c \textrm{e}^{\textrm{i}\omega_d t} \right)$ to the coupler, where $\omega_d$ is the drive frequency and $\Omega_d$ is the drive amplitude or equivalent Rabi frequency. 
Such a drive near the coupler frequency induces the ac Stark effect, which causes an energy shift of each computational state $\ket{mn0}$ according to
\begin{eqnarray}
\label{stark}
|\delta_{mn}| = \frac{1}{2} \left( \sqrt{\Delta_{mn}^2+\Omega_d^2}-|\Delta_{mn}| \right),
\end{eqnarray}
where $\Delta_{mn} = \omega_d - \omega_c^{mn}$ is the respective detuning. The sign of the shift $\delta_{mn}$ is the same as the sign of $\Delta_{mn}$.
Given the proper choice of $\omega_d$ and $\Omega_d$, the residual $ZZ$ term $\chi_{zz}^\mathrm{static}$ can be offset by the combined Stark shifts,
\begin{eqnarray}
\label{ZZ}
\chi_{zz}^\mathrm{drive} &=& \delta_{gg} + \delta_{ee} - \delta_{ge} - \delta_{eg} = -\chi_{zz}^\mathrm{static},
\end{eqnarray}
giving zero net $ZZ$ coupling.

\begin{figure*}
    \includegraphics{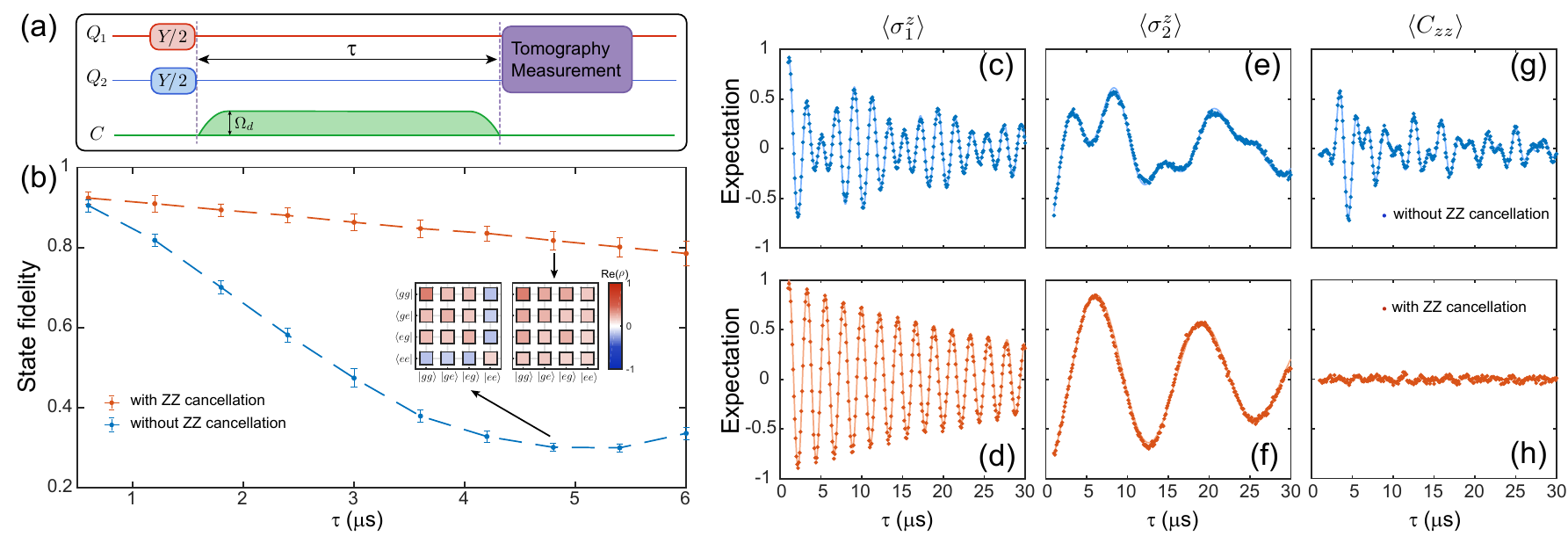}
    \caption{Verification of $ZZ$ cancellation. 
    \textbf{(a)} Sequence for measuring the two-qubit entangling phases and $ZZ$ correlations. 
    \textbf{(b)} State fidelity of the reconstructed density matrices as a function of the idling duration $\tau$ with and without the $ZZ$ cancellation drive, where error bars are the standard deviations from 14 repeated experiments. Insets are the real parts of the density matrices at $\tau=4.8~\mu$s. 
    \textbf{(c-h)} Expectation values (dots) of different observables including $\left\langle \sigma_1^z \right \rangle$~(c, d), $\left\langle \sigma_2^z \right \rangle$~(e, f), and $\left\langle C_{zz} \right \rangle$~(g, h) without (c, e, g) and with (d, f, h) the cancellation drive, measured from a simultaneous Ramsey experiment on both qubits. 
    Solid lines are from numerical simulations.}
    \label{fig3}
\end{figure*}

In our experimental device, two fixed-frequency transmon qubits ($\omega_1/2\pi=5.627$~GHz and $\omega_2/2\pi=4.353$~GHz) couple to a common 3D cavity ($\omega_c/2\pi=6.363$~GHz, $\eta_c/2\pi=-123$~kHz), a nontunable coupler, with coupling strengths $g_{1c}/2\pi = 119$~MHz and $g_{2c}/2\pi = 228$~MHz, and also directly couple to each other with a direct coupling strength $g_{12}/2\pi = 16$~MHz.
The resulting static or residual $ZZ$ coupling is $\chi_{zz}^\mathrm{static}/2\pi=-103$~kHz, which is highly consistent with the estimated one ($-101.3$~kHz)~\cite{Supplement}. The state-dependent coupler frequencies are $\omega_c^{gg}/2\pi=6.363$~GHz, $\omega_c^{eg}=\omega_c^{gg}+\chi_1$, $\omega_c^{ge}=\omega_c^{gg}+\chi_2$, and $\omega_c^{ee}=\omega_c^{gg}+\chi_1+\chi_2$, where $\chi_1/2\pi=-6.79$~MHz and $\chi_2/2\pi=-4.80$~MHz are the dispersive shifts between each qubit and the coupler (see Supplemental Material~\cite{Supplement} for more details about the device parameters and experimental setup).
For such a spectral configuration, it can be easily seen from Fig.~\ref{fig1}(b) that choosing $\omega_d$ between $\omega_c^{ee}$ and $\omega_c^{eg}$ is the most efficient way to leverage the transition frequency inhomogeneity for creating a positive $\chi_{zz}^\mathrm{drive}$ to offset $\chi_{zz}^\mathrm{static}$.
Figure \ref{fig1}(c) plots the numerically simulated net $ZZ$ coupling $\chi_{zz}$ as a function of the drive frequency and amplitude, in which the optimal choice for $ZZ$ cancellation is identified (starred) according to the objective of keeping the drive as weak as possible relative to the detunings, i.e., $\Omega_d\ll|\Delta_{mn}|$, to avoid exciting the coupler.

To calibrate the cancellation drive, we follow the simulation result to fix $\omega_d = (\omega_c^{ee} + \omega_c^{eg})/2$ and measure $\chi_{zz}$ as a function of $\Omega_d$ using the pulse sequence depicted in Fig.~\ref{fig2}(a).
For the cancellation pulse, we use a relatively slow rise and fall (300~ns for each) in order to avoid adding significant excitation upon the thermal level ($\sim$1\%) to the cavity (see Supplemental Material~\cite{Supplement} for more details). 
The sequence performs a Ramsey-like experiment on the target qubit $Q_2$ but with additional $\pi$ pulses simultaneously applied to both qubits in the middle of the sequence~\cite{Chow2013, xiong2021}. In this way, the final phase of $Q_2$ encodes the entangling $ZZ$ phase accumulated during idling while the local $Z$ phase is echoed away for both qubits. 
Figure \ref{fig2}(b) shows the measured Ramsey fringes as a function of the drive amplitude $\Omega_d$. The oscillation frequency is equivalent to the net $ZZ$ coupling $\chi_{zz}$. As the drive amplitude is increased from zero (either direction), the oscillation first slows down from $\chi_{zz}^\mathrm{static}$ to be almost invisible and then comes back again, suggesting that $\chi_{zz}$ has been tuned continuously from negative to positive.
The extracted $\chi_{zz}$ from the fit is shown in Fig.~\ref{fig2}(c) and agrees well with prediction.
The identified optimal drive amplitude $\Omega_d/2\pi \approx 0.66$~MHz (the coupling-OFF point) is adopted in the cancellation drive for subsequent experiments. Note that the flattened curve around the \lq\lq{OFF}\rq\rq{} bias, on the one hand, results from the inaccurate fitting of the Ramsey oscillation frequency near the zero $ZZ$ region, and on the other hand, may be due to the breakdown of the perturbation approximation beyond the dispersive regime~\cite{Ku2020, Ansari2019}.

To validate the cancellation, we first measure the entangling phase accrued when idling both qubits.
By preparing the system in the product state $\left(\ket{gg} + \ket{ge} +\ket{eg} +\ket{ee}\right)/2$, we perform the two-qubit state tomography after idling for a variable time $\tau$ (Fig.~\ref{fig3}(a)). 
The state fidelity of the measured density matrices at different $\tau$ with and without the $ZZ$ cancellation drive are compared in Fig.~\ref{fig3}(b).
For $\tau=4.8\,\mu$s, the two-qubit entangling phase is near its maximum, which is $\chi_{zz}^\mathrm{static}\tau\approx\pi$ in the case of no cancellation. With the cancellation drive, such a conditional phase-flip error is corrected. 
The state fidelity with cancellation shows a smooth decay with idling duration, implying that the cancellation performance is stable. One set of calibrated parameters can be used for idling operation of variable length.

We also look into other relevant observables to verify the removal of two-qubit correlations by our cancellation drive. 
In a similar experiment as Fig.~\ref{fig3}(a), we set a small detuning of $-0.5$~MHz ($+0.1$~MHz) to the microwave drives on qubit $Q_1$ ($Q_2$) in order to render a fringe pattern. At a varying delay time $\tau$, we compute the ensemble average of single-qubit observables $\left\langle \sigma_1^z \right \rangle$ and $\left\langle \sigma_2^z \right \rangle$, as well as their correlation $\left\langle C_{zz} \right \rangle =  \left\langle \sigma_1^z \sigma_2^z \right \rangle - \left\langle \sigma_1^z \right \rangle  \left\langle \sigma_2^z \right \rangle$. 
Without the cancellation drive, the Ramsey fringe exhibits a beating pattern as shown in Figs.~\ref{fig3}(c,~e), which results from the mixing of precession frequencies due to the presence of nonzero $\chi_{zz}^\mathrm{static}$. These beatings are, however, gone when the cancellation drive is applied (Figs.~\ref{fig3}(d,~f)), verifying complete $ZZ$ removal.
It can also be seen that the measured two-qubit $ZZ$ correlation functions have been significantly suppressed with the cancellation drive (Figs.~\ref{fig3}(g,~h)).
All data show good agreement with numerical simulation.

\begin{figure}
    \includegraphics{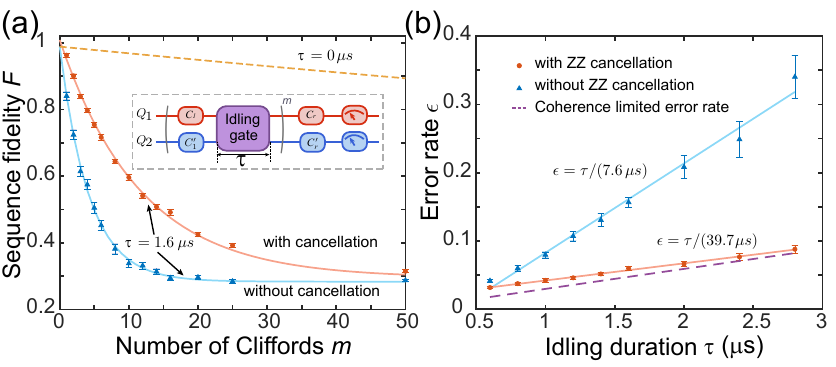}
    \caption{Randomized benchmarking of the idling gate. \textbf{(a)} Sequence fidelity $F$ as a function of the number of Clifford cycles $m$ measured by the simultaneous single-qubit RB sequence with interleaved idling gates (inset).
    Error bars are the standard deviations from the mean estimated over 80 randomized realizations. 
    \textbf{(b)} Idling gate error rate $\epsilon$ as a function of the idling duration $\tau$, obtained from fitting the decay in (a). Error bars are the estimated 95\% confidence intervals of the fitting.
    Linear fits to the data give the duration dependence of gate error $\epsilon=\tau/(7.6\,\mu\mathrm{s})$ for the case without cancellation and $\epsilon=\tau/(39.7\,\mu\mathrm{s})$ for the cases with cancellation. 
    Dashed line is the estimated coherence limited error rate.}
    \label{fig4}
\end{figure}

Finally, to benchmark the impact of residual $ZZ$ interaction on the fidelity of a general quantum circuit with intermittent idling operations and the improvement we can gain from our cancellation protocol, we implement the simultaneous Clifford-based RB experiment with interleaved idling gates~\cite{Barends2014}. 
As shown in the inset of Fig.~\ref{fig4}(a), the circuit performs simultaneous single-qubit RB with an idling gate of variable length $\tau$ inserted in each Clifford cycle.
Figure \ref{fig4}(a) shows an example of the measured sequence fidelity $F$ versus the number of Clifford cycles $m$ averaged over 80 randomizations for the case of $\tau=1.6~\mu$s. 
Comparing the interleaved case with the reference case ($\tau=0$), we obtain the idling gate error rate $\epsilon =  \frac{3}{4}(1 - p_\mathrm{int}/p_\mathrm{ref})$, where $p_\mathrm{int}$ and $p_\mathrm{ref}$ are from fitting the corresponding fidelity decay curve to $F = A p^m + B$. 
Figure~\ref{fig4}(b) compares the idling gate error rate with and without the cancellation drive at various idling gate durations $\tau$. It clearly shows significant error mitigation using the $ZZ$ cancellation drive across the entire range of $\tau$ up to $2.8~\mu$s. 
In the case without cancellation, the gate error rate has a strong $\tau$-dependence, $\epsilon=\tau/(7.6\,\mu\mathrm{s})$, which implies that, even without any active gate operation, the circuit fidelity can drop rapidly with extended idling periods. After adding the cancellation drive, the $\tau$-dependence is reduced by about 5 times to $\epsilon=\tau/(39.7\,\mu\mathrm{s})$, which agrees with the coherence limit $\epsilon \approx \frac{1}{3}(\tau/T_1^{Q_1})+\frac{1}{3}(\tau/T_1^{Q_2})$~\cite{OMalley2015}, with $T_1^{Q_1}$ and $T_1^{Q_2}$ the energy relaxation times of qubits $Q_1$ and $Q_2$, respectively. Such a full-scale improvement demonstrates the effectiveness of our cancellation protocol which removes almost all errors caused by the residual $ZZ$ interaction.

To summarize, we discuss the pros and cons of our method. 
First, our method is truly scalable given the one-to-one proportionality between the number of drives and number of connections. This is a clear advantage over previous methods based on directly driving the qubits. Second, the cancellation drive is noninvasive to the qubits because the drive frequency -- near coupler frequency -- is in a distinct band from the qubits. 
Third, the drive also causes negligible perturbation to the coupler state as it requires only a weak drive -- in our demonstrated case, an equivalent Rabi frequency of 0.66~MHz -- given a proper choice of drive parameters.
Actually, adiabatic rising and falling edges as used in our pulse further prohibits excitation of the coupler.
Fourth, our method is compatible with both nontunable and tunable couplers. Since the latter requires a local control line anyway, our method should not add any additional complexity in design.
The cost of our method is the extra microwave drive that needs to be synthesized from room-temperature electronics.
We believe, however, that our method will be useful in addressing unwanted $ZZ$ crosstalk in scalable devices, complementing other methods such as dynamical decoupling.

\begin{acknowledgments}
This work was supported by the Key-Area Research and Development Program of Guangdong Province (Grant No. 2018B030326001), the Shenzhen Science and Technology Program (Grant No. RCYX20210706092103021), the Guangdong Basic and Applied Basic Research Foundation (Grant No. 2022A1515010324), the National Natural Science Foundation of China (Grant No. 11904158), the Guangdong Provincial Key Laboratory (Grant No. 2019B121203002), the Guangdong Innovative and Entrepreneurial Research Team Program (Grant No. 2016ZT06D348), the Science, Technology and Innovation Commission of Shenzhen Municipality (Grants No. JCYJ20170412152620376, No. KYTDPT20181011104202253), and Shenzhen-Hong Kong cooperation zone for technology and innovation (Contract NO. HZQB-KCZYB-2020050).
\end{acknowledgments}


%

\end{document}


\title{Supplementary Material for ``Scalable method for eliminating residual $ZZ$ interaction between superconducting qubits"}

\author{Zhongchu Ni}
\thanks{These authors contributed equally to this work.}
\affiliation{Shenzhen Institute for Quantum Science and Engineering, Southern University of Science and Technology, Shenzhen 518055, China}
\affiliation{International Quantum Academy, Shenzhen 518048, China}
\affiliation{Department of Physics, Southern University of Science and Technology, Shenzhen 518055, China}
\affiliation{Guangdong Provincial Key Laboratory of Quantum Science and Engineering, Southern University of Science and Technology, Shenzhen 518055, China}

\author{Sai Li}
\thanks{These authors contributed equally to this work.}

\author{Libo Zhang}
\affiliation{Shenzhen Institute for Quantum Science and Engineering, Southern University of Science and Technology, Shenzhen 518055, China}
\affiliation{International Quantum Academy, Shenzhen 518048, China}
\affiliation{Guangdong Provincial Key Laboratory of Quantum Science and Engineering, Southern University of Science and Technology, Shenzhen 518055, China}

\author{Ji Chu}

\author{Jingjing Niu}
\author{Tongxing Yan}
\author{Xiuhao Deng}
\author{Ling Hu}
\author{Jian Li}
\author{Youpeng Zhong}
\author{Song Liu}

\author{Fei Yan}
\email{yanf7@sustech.edu.cn}

\author{Yuan Xu}
\email{xuy5@sustech.edu.cn}
\affiliation{Shenzhen Institute for Quantum Science and Engineering, Southern University of Science and Technology, Shenzhen 518055, China}
\affiliation{International Quantum Academy, Shenzhen 518048, China}
\affiliation{Guangdong Provincial Key Laboratory of Quantum Science and Engineering, Southern University of Science and Technology, Shenzhen 518055, China}

\author{Dapeng Yu}
\affiliation{Shenzhen Institute for Quantum Science and Engineering, Southern University of Science and Technology, Shenzhen 518055, China}
\affiliation{International Quantum Academy, Shenzhen 518048, China}
\affiliation{Department of Physics, Southern University of Science and Technology, Shenzhen 518055, China}
\affiliation{Guangdong Provincial Key Laboratory of Quantum Science and Engineering, Southern University of Science and Technology, Shenzhen 518055, China}

\maketitle
\vskip 0.5cm

\section{Experimental device and setup}
Our experimental device, similar to that in Ref.~\cite{Xu2020Demonstration, Li2021}, contains two fixed-frequency superconducting transmon qubits ($Q_1$ and $Q_2$) and a 3D coaxial stub cavity as the bus coupler~($C$). Each qubit $Q_1$ ($Q_2$) also couples to an additional $\lambda / 2$ stripline resonator $R_1$ ($R_2$) for dispersive readout. In this 3D superconducting quantum circuit architecture, strong coupling between the superconducting qubit and the coupler can be achieved by inserting the large antenna pad of the Josephson junction into the electromagnetic field of the 3D microwave cavity. Therefore, coupling strength larger than 100~MHz, even 200~MHz, can be easily achieved with the increased electric dipole moment coupling~\cite{Paik2011, Rigetti2012}. Main parameters of the system are summarized and listed in Table.~\ref{TableS1}. Note that the non-linearity of the cavity comes from the interaction with the non-linear superconducting transmon qubits~\cite{Nigg2012}.

\begin{figure*}[htbp]
  \centering
  \includegraphics{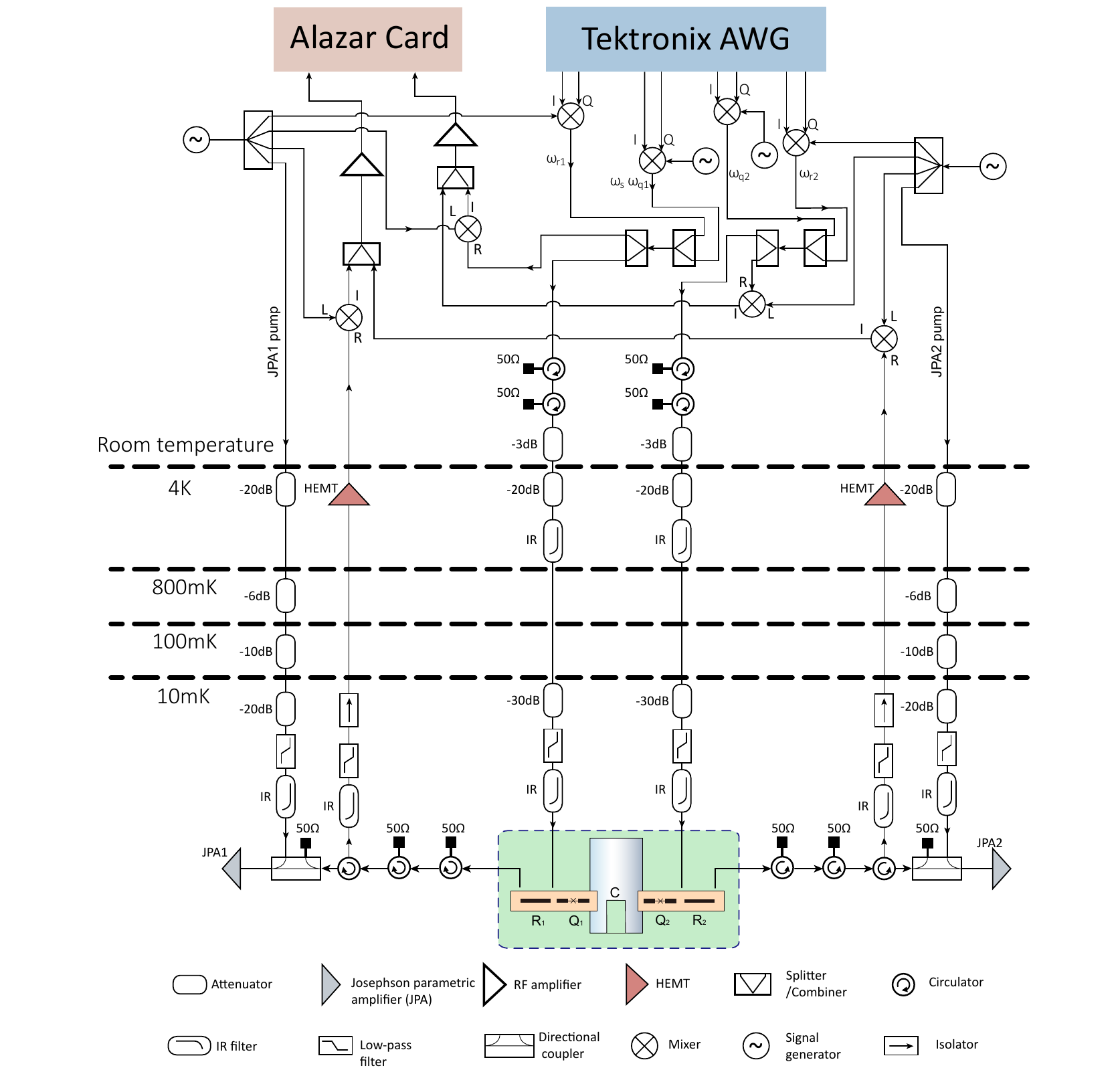}
  \caption{Details of wiring and experimental setup.}
  \label{setup}
\end{figure*}

The experimental device is installed inside a magnetic shield and cooled down to a temperature below 10~mK in a cryogen-free dilution refrigerator. Details of the experimental circuitry are shown in Fig.~\ref{setup}. The control and readout pulses of the two qubits are generated by single sideband modulations of the waveforms generated from the arbitrary waveform generator~(Tektronix 5208), as well as the control pulses for the coupler. All these control and readout signals are transmitted to the device through coaxial cables with additional isolators, filters and attenuators in order to reduce reflection waves and radiation noises. The readout signals of the two qubits are first amplified by a series of amplification chain, and finally combined and recorded by the data acquisition card~(Alazar ATS9870), as well as the corresponding reference signals. 

\begin{table}[tb]
\caption{Device parameters of the system.}
\begin{tabular}{p{6cm}<{\centering} p{1.2cm}<{\centering} p{1.2cm}<{\centering} }
  \hline
  \hline
  Parameters & $Q_1$ & $Q_2$ \\\hline
  &&\\[-0.9em]
  Readout frequency (GHz) & 8.604 & 7.997 \\
  &&\\[-0.9em]
  Qubit frequency $\omega_i/2\pi$ (GHz) & 5.627 & 4.353\\
  &&\\[-0.9em]
  Qubit anharmonicity $\eta_i/2\pi$ (MHz) & -184 & -220\\
  &&\\[-0.9em]
  $T_1$ ($\mu$s) & 18$\thicksim$22 & 20$\thicksim$28  \\
  &&\\[-0.9em]
  $T^*_2$ ($\mu$s) & 25$\thicksim$32 & 32$\thicksim$42  \\
  &&\\[-0.9em]
  $T_{\mathrm{2E}}$ ($\mu$s) & 30$\thicksim$40 & 36$\thicksim$45  \\
  &&\\[-0.9em]
  ZZ coupling strength $\chi_{zz}^\mathrm{static}/2\pi$ (kHz) & \multicolumn{2}{c}{$-103$}  \\
  &&\\[-0.9em]
  $Q$-$R$ dispersive shift $\chi_{qr}/2\pi$ (MHz) & $-2.8$ & $-1.8$  \\
  &&\\[-0.9em]
  $Q$-$C$ dispersive shift $\chi_i/2\pi$ (MHz) & $-6.79$ & $-4.80$  \\
  &&\\[-0.9em]
  Readout decay rate $\kappa_r/2\pi$ (MHz) & 1.82 & 1.32  \\
  &&\\[-0.9em]
  \hline
  \hline
  Parameters & \multicolumn{2}{c}{Coupler} \\\hline
  &&\\[-0.9em]
  Coupler frequency $\omega_c/2\pi$ (GHz) & \multicolumn{2}{c}{6.363} \\
  &&\\[-0.9em]
  Coupler anharmonicity $\eta_c/2\pi$ (kHz) & \multicolumn{2}{c}{-123} \\
  &&\\[-0.9em]
  Coupler $T_1$ ($\mu$s) & \multicolumn{2}{c}{$\thicksim$64} \\
  &&\\[-0.9em]
  Coupler $T^*_2$ ($\mu$s) & \multicolumn{2}{c}{$\thicksim$81} \\
  \hline 
  \hline
\end{tabular}
\label{TableS1}
\end{table}

\section{Theoretical calculation of the $ZZ$ cancellation}
In our system, two transmon qubits are strongly coupled to a common cavity bus coupler. The total Hamiltonian of the system can be expressed as ($\hbar \equiv 1$):
\begin{eqnarray}
\label{H}
H_0 &=& \sum_{i=1,2,c}{\omega_i \, a_i^\dag a_i + \frac{\eta_i}{2} \, a_i^\dag a_i^\dag a_i a_i} \notag\\
&+&  \sum_{i\neq j}{g_{ij} \left( a_i^\dag a_j + a_i a_j^\dag \right)},
\end{eqnarray}
where $\omega_i$ ($i\in \{ 1, 2, c \}$) is the bare frequency of the qubits or coupler, $\eta_i$ is the anharmonicity of each mode, $g_{ij}$ is their coupling strength, and $a_i$ ($a_i^\dag$) is the corresponding annihilation (creation) operator.

In the strong dispersive regime, where $g_{ij} \ll \left| \omega_i - \omega_j \right|$, we can conveniently diagonalize the above Hamiltonian to the dispersive Hamiltonian:
\begin{eqnarray}
\label{H0}
H_0 &=& \sum_{m,n=g,e}{\ket{mn}\bra{mn} \otimes \left(\omega_c^{mn} a_c^\dag a_c + \frac{\eta_c}{2} \, a_c^\dag a_c^\dag a_c a_c\right)}  \notag\\
&+ &  \chi_{zz}^\mathrm{static}\,\ket{ee}\bra{ee}\otimes\ket{0}\bra{0},
\end{eqnarray}
in the rotating frame of the two qubits after truncation into the relevant subspace. Here, $\ket{g}$ ($\ket{e}$) denotes the ground (excited) state of the qubits and $\ket{0}$, $\ket{1}$, $\ket{2}$... denote the coupler state. 
$\omega_c^{mn}$ is the 0-1 transition frequency of the coupler with the two qubits in state $\ket{mn}$.
$\eta_c$ is the anharmonicity of the coupler.
$\chi_{zz}^\mathrm{static}$ is the residual $ZZ$ coupling. The qubit-state-dependent coupler frequencies are $\omega_c^{eg} = \omega_c^{gg} + \chi_1$, $\omega_c^{ge} = \omega_c^{gg} + \chi_2$, and $\omega_c^{ee} = \omega_c^{gg} + \chi_1 + \chi_2$, respectively, where $\chi_1$ and $\chi_2$ are the dispersive shifts between the two qubits and the coupler. These dispersive shifts relate to the frequency detunings and coupling strengths between the two qubits and the coupler~\cite{koch2007transmon}. As a result, from the measured dispersive shifts $\chi_1/2\pi = -6.79$~MHz and $\chi_2/2\pi = -4.80$~MHz, and the static residual $ZZ$ coupling $\chi_{zz}^\mathrm{static}/2\pi=-103$~kHz, we could deduce the coupling strengths $g_{12}/2\pi=16$~MHz, $g_{1c}/2\pi=119$~MHz, and $g_{2c}/2\pi=228$~MHz, respectively, from numerical diagonalization of the Hamiltonian in Eq.~(\ref{H}). 

With these Hamiltonian parameters, we could also estimate the static $ZZ$ coupling by using a fourth-order perturbation theory~\cite{Chu2021}. The static ZZ coupling can be expressed as
\begin{eqnarray}
\chi_{zz}^\mathrm{static} &\approx&   \frac{2\{(\eta_1+\eta_2)\Tilde{g}^2 -2v[2\eta_1\eta_2 + (\eta_1-\eta_2)\Delta_{12} ] \Tilde{g} \} }{(\Delta_{12}+\eta_1)(\Delta_{12}-\eta_2) } \notag\\
&+ &  2v^2\left[4\eta_c + \frac{(\eta_1+\eta_2)\Delta_{12}^2}{(\Delta_{12}+\eta_1)(\Delta_{12}-\eta_2)} \right],\notag
\end{eqnarray}
where $\Tilde{g}=g_{12}+g_{1c}g_{2c}(1/\Delta_{1c}+1/\Delta_{2c}-1/\Sigma_{1c}-1/\Sigma_{2c})/2$ is the effective coupling strength between the two qubits, $v=g_{1c}g_{2c}/(2\Delta_{1c}\Delta_{2c})$ is a small dimensionless quantity in the dispersive limit, $\Delta_{ic}=\omega_i-\omega_c$ is frequency detuning between the two qubits and the coupler, and $\Sigma_{ic}=\omega_i+\omega_c$ is their frequency summation. From the device parameters given above, we could estimate the effective coupling strength between the two qubits is $\Tilde{g}/2\pi=-11.6$~MHz, and the static residual $ZZ$ coupling between the two qubits is $\chi_{zz}^\mathrm{static}/2\pi=-101.3$~kHz, which is consistent with our measured value ($-103$~kHz).

In order to cancel $\chi_{zz}^\mathrm{static}$, we apply a microwave drive to the coupler, with a drive Hamiltonian of $H_d = \frac{\Omega_d}{2} \left(e^{-i\omega_d t} a_c^\dag  + e^{i \omega_d t} a_c\right)$. Here $\Omega_d$ ($\omega_d$) is the amplitude (frequency) of this drive. In a rotating frame of the drive frequency, the Hamiltonian of the coupler with the two qubits in state $\ket{mn}$ can be expressed as:
\begin{eqnarray}
\label{H1}
H_{mn} &=& -\Delta_{mn} a_c^\dag a_c + \frac{\eta_c}{2}a_c^\dag a_c^\dag a_c a_c \notag\\
&+& \frac{\Omega_d}{2} \left(a_c^\dag + a_c\right),
\end{eqnarray}
where $\Delta_{mn}=\omega_d-\omega^{mn}_c$ is the frequency detuning between the microwave drive and the coupler frequency with the two qubits in $\ket{mn}$ state. With weak drive to the coupler, the above Hamiltonian can be approximately truncated into the \{\ket{0},\ket{1}\} subspace with 
\begin{eqnarray}
H_{mn} = \begin{pmatrix}
0 & \Omega_d/2 \\
\Omega_d/2 & -\Delta_{mn}
\end{pmatrix}.
\label{eqs3} 
\end{eqnarray}
and can be diagonalized with the dressed eigenstates 
\begin{eqnarray}
\ket{\psi_{mn}^{\pm}} &=& \frac{\Omega_d}{\sqrt{2\Delta_{mn}^2+2\Omega_d^2\mp2\Delta_{mn}\sqrt{\Delta_{mn}^2+\Omega_d^2}}}\ket{0} \notag\\
&+& \frac{-\Delta_{mn}\pm\sqrt{\Delta_{mn}^2+\Omega_d^2}}{\sqrt{2\Delta_{mn}^2+2\Omega_{d}^2\mp2\Delta_{mn}\sqrt{\Delta_{mn}^2+\Omega_d^2}}}\ket{1} \notag\\
\end{eqnarray}
and eigenvalues 
\begin{eqnarray}
E_{mn}^{\pm} = \frac{-\Delta_{mn}\pm\sqrt{\Delta_{mn}^2+\Omega_d^2}}{2},
\label{eqs5} 
\end{eqnarray}
respectively.

In the limit $\Omega_d \ll |\Delta_{mn}|$, the eigenstate of the coupler still remains in the vacuum state. Thus, the ac Stark shift of the state $\ket{mn0}$ can be expressed as:
\begin{eqnarray}
|\delta_{mn}| = \frac{1}{2} \left( \sqrt{\Delta_{mn}^2+\Omega_d^2}-|\Delta_{mn}| \right),
\label{eqs6} 
\end{eqnarray}
and the sign of $\delta_{mn}$ is the same as the sign of $\Delta_{mn}$. Due to the inhomogeneity of the coupler frequency $\omega_c^{mn}$ for different two-qubit states, this drive will induce different frequency shifts for the two-qubit states. By precisely engineering these frequency shifts, we could achieve
\begin{eqnarray}
\chi_{zz} & = & \chi_{zz}^\mathrm{drive} + \chi_{zz}^\mathrm{static} \\\notag
& = & \delta_{gg}+\delta_{ee} - \delta_{ge}-\delta_{eg} + \chi_{zz}^\mathrm{static} \\\notag
& = & 0,
\end{eqnarray}
to completely cancel the static $ZZ$ coupling term and make the total $ZZ$ coupling strength equalling zero. Our method is also applicable to tunable coupler with frequency-tunable qubit due to no constrain on the coupler anharmonicity. The $ZZ$ coupling can also be enhanced to realize two-qubit entangling gate with increasing the drive amplitude, which is similar to the resonator-induced phase gate~\cite{Paik2016}. Note that when applying a detuned drive on the qubit, we could also achieve photon-number-dependent frequency shift of the coupler states, which is similar to the photon-number-resolved ac Stark shift technique for bosonic mode~\cite{Ma2020}.

We perform the numerical simulations of our $ZZ$ cancellation approach by directly diagonalizing the total Hamiltonian and calculate the net $ZZ$ coupling as a function of the drive frequency and amplitude, with the results shown in Fig.1(d) in the main text. By choosing the drive frequency $\omega_d = \left(\omega_c^{eg}+\omega_c^{ee}\right)/2$, we could achieve continuous adjustment of the net $ZZ$ coupling from negative to positive by varying the drive amplitude $\Omega_d$. The optimal parameters for $ZZ$ cancellation is chosen according to the objective of keeping the drive as weak as possible relative to detunings -- i.e.\ $\Omega_d\ll|\Delta_{mn}|$ -- to avoid exciting the coupler.

\section{Measurement of the $ZZ$ coupling}
In the main text, we have demonstrated that the net ZZ coupling can be continuously tuned with varying the pulse amplitude of the cancellation drive on the coupler, with the experimental results shown in Fig.~2 in the main text. Note that the measured $\chi_{zz}$ at the "OFF" points in this figure are estimated to be less than 8~kHz, limited by the total evolution duration of 30 $\mu$s. Here we also perform another experiment to further verify the $ZZ$ cancellation of our control approach.

\begin{figure}[bth]
  \centering
  \includegraphics{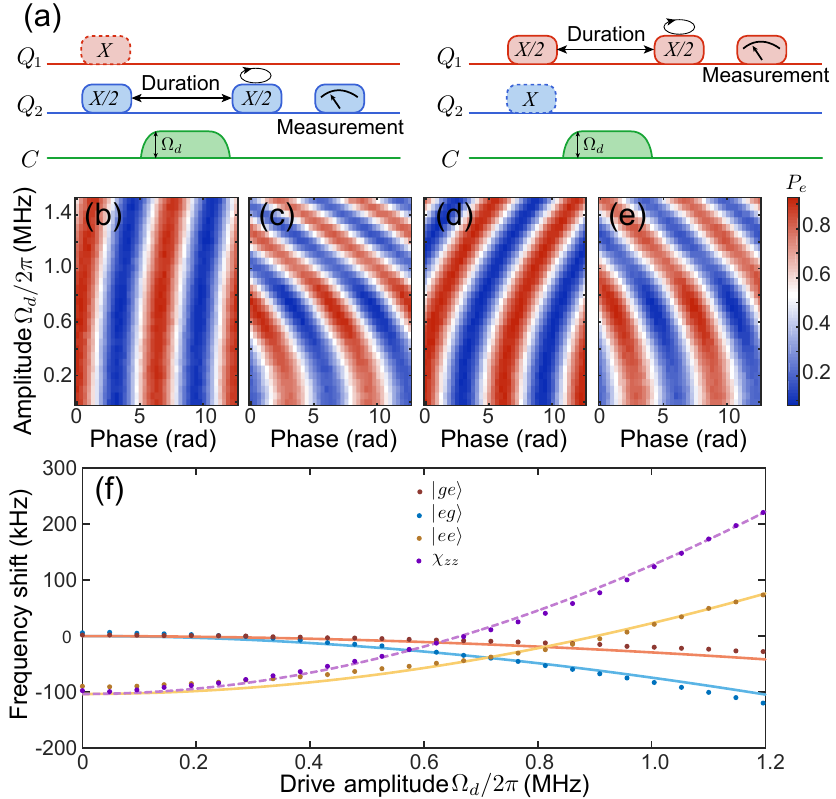}
  \caption{Measurement of the $ZZ$ coupling with conditional Ramsey experiment. (a) Experimental pulse sequences. (b-e) Measured Ramsey fringes of qubit $Q_2$~(b, c) and $Q_1$~(d, e) as a function of the drive amplitude, with the other qubit initially prepared in ground~(b, d) and excited~(c, e) state, respectively. (f) Frequency shifts of the two-qubit states $\ket{ge}$, $\ket{eg}$, and $\ket{ee}$, with relative to $\ket{gg}$, and the total $ZZ$ coupling as a function of the drive amplitude. The symbols and lines are experimental data and numerical simulations, respectively.}
  \label{fig2}
\end{figure}

In this experiment, the two-qubit $ZZ$ coupling is measured through a conditional Ramsey experiment, with the experimental sequence shown in Fig.~\ref{fig2}(a). Here, we first prepare one qubit in ground or excited state, and then perform a Ramsey experiment on the other qubit to measure the accumulated phase shift during the fixed idling period~($\tau = $5~$\mu$s), with varying the rotation phases of the second $\pi/2$ pulse in the Ramsey experiment. The cancellation drive is applied during the idling period with a variable drive amplitude.

The experimental results are shown in Figs.~\ref{fig2}(b-e). From fitting the Ramsey oscillations, we could extract the accumulated phase shifts $\Delta \Phi$ during the idling period for the transitions of $\ket{gg}\leftrightarrow\ket{ge}$, $\ket{eg}\leftrightarrow\ket{ee}$, $\ket{gg}\leftrightarrow\ket{eg}$, and $\ket{ge}\leftrightarrow\ket{ee}$, respectively. Thus the corresponding transition frequencies $E_{ge} - E_{gg}$, $E_{ee} - E_{eg}$, $E_{eg} - E_{gg}$, and $E_{ee} - E_{ge}$ can be extracted by $\Delta \Phi/\tau$. Here $E_{mn}$ denotes the energy level for the two qubits in state $\ket{mn}$ and the coupler in vacuum state $\ket{0}$. Figure~\ref{fig2}(f) shows the frequency shifts as a function of the drive amplitude $\Omega_d$, for the two-qubit state $\ket{ge}$, $\ket{eg}$, and $\ket{ee}$, relative to the $\ket{gg}$ state. The calculated net $ZZ$ coupling $\chi_{zz} = E_{gg} + E_{ee} - E_{ge} - E_{eg}$ is also consistent with that in Fig.~2 in the main text. With the cancellation drive, the two qubits also have small frequency shifts during the idling period, which are compensated by applying corresponding small detunings to the microwave drives on the two qubits in the following experiments.

\section{Measuring the two-qubit entangling phases}
To validate the cancellation, we first measure the entangling phase accrued when idling both qubits, with the experimental results shown in Fig.~3(b) in the main text. In this experiment, we first initialize the two qubits in a product state of $\left(\ket{gg}+\ket{ge}+\ket{eg}+\ket{ee}\right)/2$, apply a two-qubit idling gate with different durations, and finally perform two-qubit state tomography measurement~\cite{Nielsen} of the final state. The density matrix of the final state is reconstructed by 16 pre-rotations $U_k \in \{I, X/2, Y/2, X \}^{\otimes 2}$ before measurements, giving a result of $\langle M_k \rangle = \mathrm{Tr}\left(\rho U_k^{\dagger}M_I U_k\right)$ with $k=1,2...16$. Here $M_I = \ket{gg}\bra{gg}$ is the two-qubit measurement operator with projecting on the two-qubit ground state. The density matrix of the final state of the two qubits is reconstructed by the maximum likehood estimation method~\cite{James2001} with these measured observables $\langle M_k \rangle$. The state fidelity is defined as $F = \left| \mathrm{Tr} \left(\rho_\mathrm{exp} \rho_\mathrm{ideal} \right) \right|$, and shown in Fig.~3(a) in the main text as a function of the idling gate duration with and without the cancellation drive. Here $\rho_\mathrm{exp}$ and $\rho_\mathrm{ideal}$ represent the experimentally measured and ideal density matrices of the two-qubit state.

\begin{figure}[tb]
  \centering
  \includegraphics{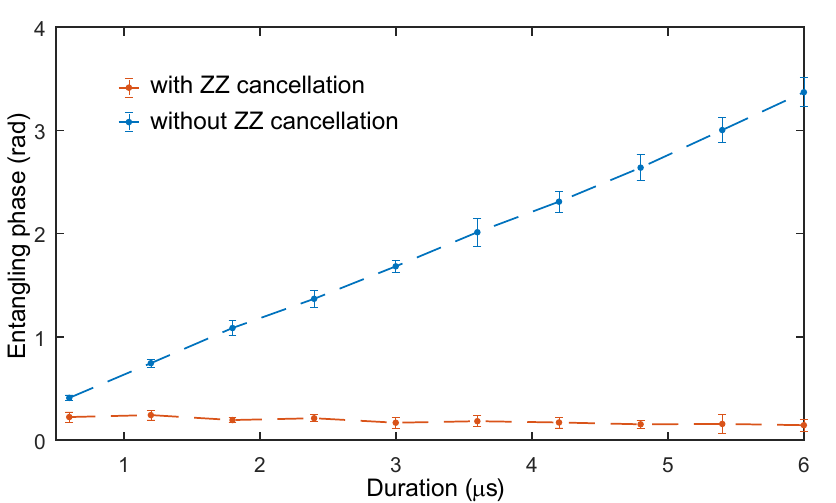}
  \caption{Measured two-qubit entangling phases as a function of the idling gate duration with and without the $ZZ$ cancellation drive. 
  }
  \label{entangling_phase}
\end{figure}

Here in Fig.~\ref{entangling_phase}, we also present the two-qubit entangling phase $\Delta \phi = \phi_{ee} - \phi_{ge} - \phi_{eg}$ of the final state as a function of the idling gate duration $\tau$ for the two cases. Here $\phi_{ge}$, $\phi_{eg}$, and $\phi_{ee}$ are the phases of the final state for the corresponding two-qubit basis states relative to the $\ket{gg}$ state. Obviously, with the $ZZ$ cancellation, the two-qubit entangling phase has been greatly suppressed due to the complete cancellation of the two-qubit static $ZZ$ coupling. Note that there still exist small entangling phases with cancellation, which is due to the phase accumulations during the period of the rising and falling edges of the cancellation pulse.

\section{Measuring the two-qubit $ZZ$ correlations}
\begin{figure}[tb]
  \centering
  \includegraphics{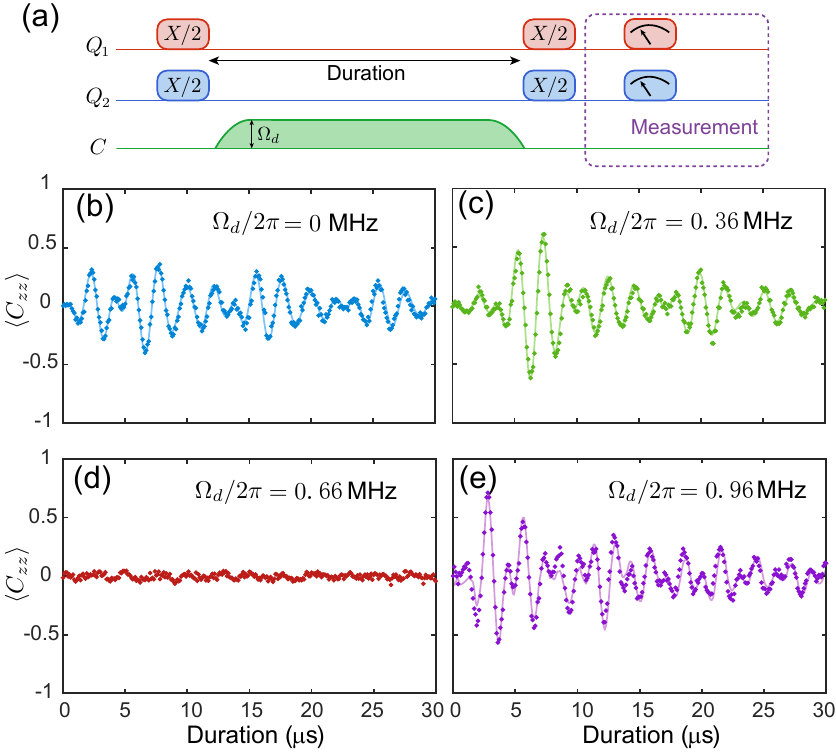}
  \caption{Measurement of the two-qubit $ZZ$ correlations. (a) Experimental pulse sequence. (b-e) Measured two-qubit $ZZ$ correlations with the amplitude of the $ZZ$ cancellation drive $\Omega_d/2\pi = 0$, 0.36, 0.66, and 0.96~MHz, respectively. The experimental data (symbols) agree well with the numerical simulations (solid lines).}
  \label{correlation}
\end{figure}

In the presence of the $ZZ$ cancellation drive, we also demonstrate the suppression of the two-qubit $ZZ$ correlations by a two-qubit simultaneous Ramsey experiment, with the experimental results shown in Fig.~3(c-h) in the main text. The corresponding experimental pulse sequence is depicted in Fig.~\ref{correlation}(a), where both qubits are simultaneously performed a pair of $\pi/2$ pulses with different idling durations between them, and finally measured by simultaneous single-shot readout of the two qubits. From the measured two-qubit populations of $P_{gg}$, $P_{ge}$, $P_{eg}$, and $P_{ee}$, we could calculate the ensemble average of the observables $\left\langle \sigma_1^z \right \rangle$ and $\left\langle \sigma_2^z \right \rangle$ of qubit $Q_1$ and $Q_2$, respectively, as well as the correlation $\left\langle C_{zz} \right \rangle =  \left\langle \sigma_1^z \sigma_2^z \right \rangle - \left\langle \sigma_1^z \right \rangle  \left\langle \sigma_2^z \right \rangle$ between these two qubits. In order to render a fringe pattern, we add a small detuning of $-0.5$~MHz and $+0.1$~MHz to the microwave drives on qubit $Q_1$ and $Q_2$, respectively. The corresponding experimental results are shown in Fig.~3(c-h) in the main text.

Here we also present the experimental data of the two-qubit $ZZ$ correlations when applying the $ZZ$ cancellation drive with different amplitudes. And the corresponding experimental results are shown in Fig.~\ref{correlation}(b-e) for the drive amplitude $\Omega_d/2\pi= 0$, $0.36$, $0.66$, and $0.96$~MHz, respectively. Obviously, only when the drive amplitude of the microwave tone equals to the value to completely cancel the static $ZZ$ coupling, the oscillation beatings have been completely removed and the $ZZ$ correlations have been greatly suppressed. Note that for these data, the detunings applied on the two qubits are $-0.5$~MHz and $-0.1$~MHz, respectively.

\begin{table}[b]
    \centering
	\renewcommand\arraystretch{1.5}
	\caption{The 24 single-qubit Clifford gates written in terms of physical qubit gates \{$I$, $\pm X$, $\pm Y$, $\pm X/2$, $\pm Y/2$\} and virtual Z gates \{$\pm Z$, $\pm Z/2$\}. The gates are applied in time order. }
	\begin{tabular}{p{3.5cm}<{\centering}p{4.8cm}}
		\hline
		\hline
		\multirow{4}{*}{Paulis} & \  \  \  $I$ \\
		&\  \  $X$ \\
		& \  \  $Y$ \\
		& \  \  $Z$,\quad\quad\quad\quad\quad\  \  $I$ \\\hline
		\multirow{8}{*}{$2\pi/3$ rotations} & \  \  $Y/2$,\quad\quad\quad\  \  $-Z/2$\\
		& $-Y/2$,\quad\quad\quad\  \  \  \  $Z/2$\\
		& \  \  $Y/2$,\quad\quad\quad\  \  \  \  $Z/2$\\
		& $-Y/2$,\quad\quad\quad\  $-Z/2$\\
		& \  \  $X/2$,\quad\quad\quad\  $-Z/2$\\
		& $-X/2$,\quad\quad\quad\  \  \  $Z/2$\\
		& \  \  $X/2$,\quad\quad\quad\  \  \  $Z/2$\\
		& $-X/2$,\quad\quad\quad$-Z/2$\\\hline
		\multirow{6}{*}{$\pi/2$ rotations} & \  \  $X/2$\\
		& $-X/2$\\
		& \  \  $Y/2$\\
		& $-Y/2$\\
		& \  \  $Z/2$,\quad\quad\quad\  \  \; $I$\\
		& $-Z/2$,\quad\quad\quad\  \  \  $I$\\\hline
		\multirow{6}{*}{Hadmard-like} & \  \  $Y/2$,\quad\quad\quad\;$-Z$\\
		& $-Y/2$,\quad\quad\quad\quad  $Z$\\
		& \  \  $X/2$,\quad\quad\quad\quad  $Z$\\
		& $-X/2$,\quad\quad\quad $-Z$\\
		& $-X$,\quad\quad\quad\quad\  \;$Z/2$\\
		& $-Z/2$,\quad\quad\quad $-Y$\\
		\hline
		\hline
	\end{tabular} \vspace{-6pt}
	\label{tabs2}
\end{table}

\section{Randomized benchmarking}
In the main text, the performance of the two-qubit idling gate is characterized with simultaneous Clifford-based randomized benchmarking~(RB) method. In the RB experiment, we first perform $m$ random single-qubit Clifford gates for both qubit $Q_1$ and $Q_2$, insert a two-qubit idling gate with a duration of $\tau$ after each Clifford gate, append an additional recovery gate on the two qubits to inverse the whole sequence, and finally measure the two qubits simultaneously. The single-qubit Clifford gate set is constructed with physical gates \{$I$, $\pm X$, $\pm Y$, $\pm X/2$, $\pm Y/2$\} and virtual Z gates~\cite{McKay2017} \{$\pm Z$, $\pm Z/2$\}, respectively, and listed in Table.~\ref{tabs2}. This construction ensures that each Clifford gate has the same gate length. From fitting the measured sequence fidelity decay curves with an equation of $F = Ap^m+B$, we could obtain the decay constants $p_\mathrm{ref}$ and $p_\mathrm{int}$ for the cases without and with interleaved idling gate, respectively. Thus, the two-qubit idling gate error rate can be expressed as $\epsilon = \frac{3}{4}(1 - p_\mathrm{int}/p_\mathrm{ref})$. The experimentally measured error rate as a function of the idling gate duration is shown in Fig.~4(b) in the main text. Coherence limited fidelity for the idling gate with a duration of $\tau$ for qubit $Q_1(Q_2)$ is estimated by $F_{1(2)} = 1-\frac{1}{3}(\tau/T_1^{Q_1(Q_2)})$, with $T_1^{Q_1}$ and $T_1^{Q_2}$ the energy relaxation times for qubit $Q_1$ and $Q_2$, respectively. Here, the factor $"1/3"$ denotes that one-excitation loss error gives an average infidelity of $1/3$ during the Clifford-based RB experiment~\cite{OMalley2015}. Therefore, coherence limit error rate for two-qubit idling gate is estimated by $\epsilon = 1-F_1\cdot F_2\approx \frac{1}{3}(\tau/T_1^{Q_1})+\frac{1}{3}(\tau/T_1^{Q_2})$ and plotted in Fig.~4(b) in the main text.

\section{Error analysis}
The detuned microwave drive on the coupler will induce frequency shifts of the two-qubit states to cancel the static $ZZ$ coupling, but also can excite the coupler during the operation due to the off-resonant driving. This coupler leakages should be minimized in order to avoid harmful effect on the two qubits. This is achieved by optimizing the drive frequency and amplitude to simultaneously realize the cancellation of $ZZ$ coupling and the minimization of the coupler excitation. The optimized parameters are marked with a star marker in Fig.~1(d) in the main text.

\begin{figure}[b]
  \centering
  \includegraphics{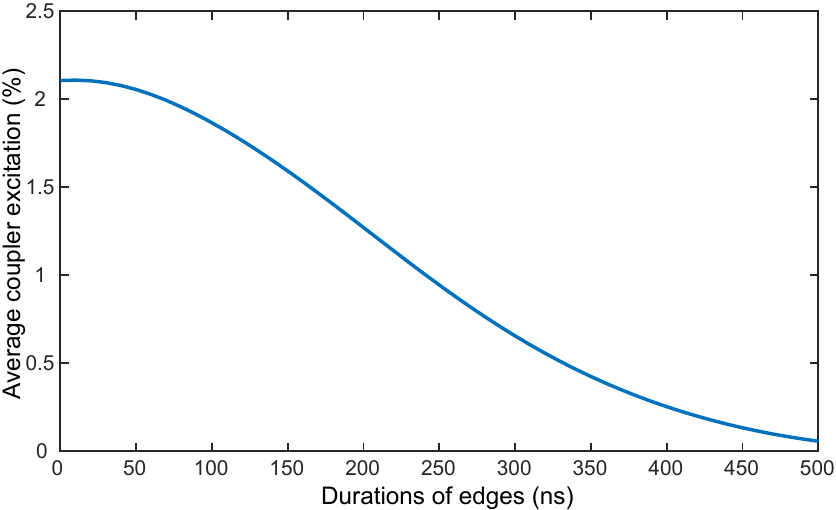}
  \caption{Average coupler excitations after the $ZZ$ cancellation drive as a function of the durations of the rising and falling edges for the detuned microwave drive on the coupler.}
  \label{leakage}
\end{figure}

In order to further suppress the coupler leakage after the idling operation, we turn on the cancellation drive adiabatically with smooth rising and falling edges. Here we simulate the average coupler excitations as a function of the duration of the rising and falling edges, with the simulation results shown in Fig.~\ref{leakage}. The drive amplitude and frequency of the $ZZ$ cancellation tone are the same with that used in our experiment. Obviously, when increasing the duration of rising and falling edges, the coupler excitations could be further minimized due to the more adiabatic of the cancellation pulse on the coupler. In our experiment, we choose a reasonable value of $300$~ns for the rising and falling edges of the cancellation drive thus making the average coupler excitations suppressed to the typical thermal excitations~($\sim$1\%) of the two qubits and coupler. 

\begin{figure}[tbh]
  \centering
  \includegraphics{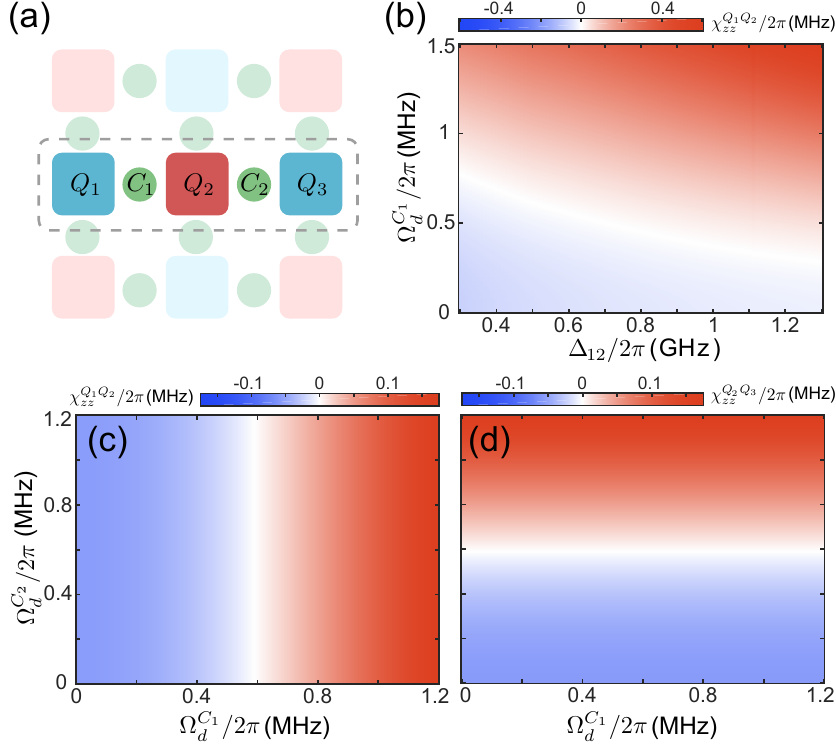}
  \caption{(a) Layout of a 2D qubit array in a square lattice, where the red and blue squares represent fixed-frequency superconducting qubits in different frequency bands and the green circulars are for tunable couplers. Here we mainly focus on the one dimensional array with three qubits and two couplers, as indicated by the gray dashed box. The corresponding device parameters are chosen as: $(\omega_{Q_1}, \omega_{Q_2}, \omega_{Q_3})/2\pi = (5.1, 5.6, 5.0)$ GHz, $(\eta_{Q_1}, \eta_{Q_2}, \eta_{Q_3})/2\pi = (-0.2,-0.2,-0.2)$ GHz, $(\omega^\mathrm{idle}_{C_1}, \omega^\mathrm{idle}_{C_2})/2\pi = (6.5, 6.4)$ GHz, $(\eta_{C_1}, \eta_{C_2})/2\pi = (-0.3,-0.3)$ GHz, $(g_{Q_1Q_2}, g_{Q_1C_1}, g_{Q_2C_1})/2\pi = (10, 120, 80)$ MHz, and $(g_{Q_2Q_3}, g_{Q_2C_2}, g_{Q_3C_2})/2\pi = (10, 70, 120)$ MHz for the following numerical simulations. (b) The simulated total $ZZ$ coupling $\chi_{zz}^{Q_1Q_2}$ as a function of the qubit detuning $\Delta_{12}$ between qubits $Q_1$ and $Q_2$, and the amplitude $\Omega_d^{C_1}$ of the cancellation drive on coupler $C_1$. (c, d) The simulated total $ZZ$ couplings $\chi_{zz}^{Q_1Q_2}$~(c) and $\chi_{zz}^{Q_2Q_3}$~(d) as a function of the amplitudes $\Omega_d^{C_1}$ and $\Omega_d^{C_2}$ of the cancellation drives on couplers $C_1$ and $C_2$.}
  \label{scalability}
\end{figure}

\section{Scalability of the $ZZ$ cancellation method}
In this section, we discuss the scalability of our $ZZ$ cancellation method. Considering the tunable-coupling architecture with fixed-frequency superconducting qubits, we present a 2D qubit array in a square lattice, with the layout shown in Fig.~\ref{scalability}(a). In this layout, the red and blue squares representing fixed-frequency superconducting qubits in different frequency bands are alternately arranged to keep a frequency detuning of about 500 MHz between neighboring qubits. Their direct coupling strength is on the order of 10 MHz, while the coupling strengths between each qubit and the adjacent couplers are about 100 MHz. With this choice of device design, the residual $ZZ$ couplings between neighboring qubits can be made well below 100 kHz, and could be further cancelled by simultaneously applying independent cancellation drives on all the couplers.

Besides, the frequency detuning between neighboring qubits has a large degree of freedoms for device design. We perform numerical simulations of the total $ZZ$ coupling $\chi_{zz}^{Q_1Q_2}$ between qubits $Q_1$ and $Q_2$, as a function of their frequency detuning $\Delta_{12}$ and the cancellation drive amplitude $\Omega_d^{C_1}$ on coupler $C_1$. The result shown in Fig.~\ref{scalability}(b), indicates that we can always find an appropriate cancellation drive amplitude to zero the total $ZZ$ coupling for a large range of frequency detunings between neighboring qubits, thus further verifying the high scalability of our $ZZ$ cancellation method in future large-scale quantum processor.

Our $ZZ$ cancellation method allows for independent addressability of each neighboring qubit-qubit connection, and thus can be applied simultaneously on all couplers. Considering an example of a linear array with 3 qubits and 2 couplers, we perform numerical simulations of the simultaneous $ZZ$ cancellation for the two neighboring connections. The simulated total $ZZ$ couplings $\chi_{zz}^{Q_1Q_2}$ and $\chi_{zz}^{Q_2Q_3}$ as a function of the cancellation drive amplitudes $\Omega_d^{C_1}$ and $\Omega_d^{C_2}$ are shown in Figs.~\ref{scalability}(c, d). These simulation results indicate that the $ZZ$ cancellation of one qubit-qubit pair does not affect that of other adjacent qubit-qubit pairs, thus further verifying that our $ZZ$ cancellation method works independently and can be conveniently extend to a larger qubit array.

%